\documentclass[a4paper,twocolumn,10pt,accepted=2025-07-18]{quantumarticle}

\pdfoutput=1
\usepackage[utf8]{inputenc}
\usepackage[english]{babel}
\usepackage[T1]{fontenc}
\usepackage{amsmath}
\usepackage{hyperref}
\usepackage[numbers]{natbib}

\usepackage{etoolbox}
\usepackage{amssymb}
\usepackage{enumitem}
\usepackage{leftindex}

\newcommand{\ket}[1]{\left | #1 \right \rangle}
\newcommand{\ketNoResize}[1]{ | #1 \rangle}
\newcommand{\bra}[1]{\left \langle #1 \right |}
\newcommand{\braket}[2]{\left \langle #1 | #2 \right \rangle}
\newcommand{\beq}{\begin{equation}}
\newcommand{\eeq}{\end{equation}}
\newcommand{\ra}{\rangle}
\newcommand{\la}{\langle}

\newcommand{\hL}{{\bf{\hat L}}}
\newcommand{\hq}{{\bf{\hat q}}}
\newcommand{\htheta}{{\bf{\hat \theta}}}

\begin{document}

\title{Conservation Laws For Every Quantum Measurement Outcome}

\author[1]{Daniel Collins}
\orcid{0009-0009-0365-4206}
\author[1]{Sandu Popescu}
\orcid{0000-0003-1142-7694}
\affil[1]{H. H. Wills Physics Laboratory, University of
Bristol, Tyndall Avenue, Bristol BS8 1TL}

\date{Jul 2025}

\begin{abstract}

In the paradigmatic example of quantum measurements, whenever one measures a system which starts in a superposition of two states of a conserved quantity, it jumps to one of the two states, implying different final values for the quantity that should have been conserved.  The standard law of conservation for quantum mechanics handles this jump by stating only that the total {\it distribution} of the conserved quantity over repeated measurements is unchanged, but states nothing about individual cases.  Here however we show that one can go beyond this and have conservation in each individual instance. We made our arguments in the case of angular momentum of a particle on a circle, where many technicalities simplify, and bring arguments to show that this holds in full generality. Hence we argue that the conservation law in quantum mechanics should be rewritten, to go beyond its hitherto statistical formulation, to state that the total of a conserved quantity is unchanged in every individual measurement outcome. As a further crucial element, we show that conservation can be localised at the level of the system of interest and its relevant frame of reference, and is independent of any assumptions on the distribution of the conserved quantity over the entire universe. 

\end{abstract}

\maketitle

\section{Introduction}

Conservation Laws and their counterpart, symmetries, permeate our view of physics.  They arise at the most fundamental levels, such as rotational symmetry leading to conservation of angular momentum, are useful at a practical level, such as conservation of momentum helping us calculate the outcome of a particle collision, and can even be used to help us derive theories. At a general level, classical conservation laws state that the sum total across all systems of a conserved quantity (momentum, angular momentum, charge, etc.) remains unchanged over the course of an experiment.  However, in quantum mechanics a state can be in a superposition of different values of a conserved quantity.  If we then measure the conserved quantity, we get one of the values it had in the superposition: It seems that the conserved quantity, which is supposed to be unchanging, jumps to the new value upon measurement. The standard way quantum mechanics deals with this problem is to say that what conservation laws mean is that if we repeat measurements on a statistical ensemble of identically prepared systems, and measure the conserved quantity at the start of each experiment or, alternatively, at the end, we get the same {\it statistical distribution}. 

Whilst this conservation of distribution law has been used successfully for a century, conceptually it is troubling.  It still does not explain why the conserved quantity, which is supposed to be unchanging, jumps to the new value upon measurement. 

In classical physics this could be easily explained, by thinking that we have dealt with a statistical ensemble all along: each system initially had a different value of the conserved quantity, only that we didn't know it, and the measurement at the end simply revealed it. The "jump" is therefore apparent - nothing happens in Nature and it is only a jump in our knowledge. But an attempt of thinking along these lines in quantum mechanics is tantamount to thinking of a hidden variables model which we know doesn't work. The problem therefore holds. 

This is one of the most fundamental problems in quantum mechanics, however it is rarely discussed in the literature \cite{manyWorldsConservation}, and whilst a number of solutions have been proposed, none of them, in our opinion, are satisfactory. 

In the present paper however we argue that there actually 
{\it is} conservation for each individual outcome: we show that the
jump discussed before is simply an optical illusion, and
that a more careful analysis reveals the conservation. 

The key to our results follows from a significant breakthrough that came in \cite{boxParadoxQuestion,boxParadoxSolution}, which opened the way of considering conservation in individual cases.  In \cite{boxParadoxQuestion} a situation was presented in which the usual explanation that in a measurement of a conserved quantity ``we simply found what it was there'' doesn't work and therefore showed that the issue  of conservation in individual instances can not be ignored and has to be considered seriously. Specifically, a situation was presented in which a particle was prepared in a box with a superposition of low energies. Yet, when the box is opened in a particular place for a short time, the particle sometimes emerges from the box with a high energy, much higher than the maximal energy it initially had.  One might have thought that this high energy came from the device used to open the box, but it was proven that the energy of the opener did not change.  Statistically, looking at the cases where the particle emerged from the box and the cases when the particle remained inside, the total energy (particle plus opener) is conserved.  Hence relying only on the standard, statistical conservation law would have led one to ignore this situation.  Arguing however that the high energy must have come from somewhere forced one to look further, and finally a solution was found  \cite{boxParadoxSolution}.  A new concept was introduced: the need to look not only how the system is measured, but also to look {\it how it was prepared}. Usually the way in which a state is prepared is ignored - indeed,  once the initial state is known, that's all we need: how that was prepared makes no difference. Yet, as argued in \cite{boxParadoxSolution} the preparation involves a frame of reference, and while for most questions we don't need to explicitly refer to it, {\it not even for the usual statistical conservation}, to extend conservation to individual cases the frame of reference is essential. 

In the present paper we go much further by going from the single, particular, very special situation presented in \cite{boxParadoxQuestion,boxParadoxSolution} to a very large and natural class of situations, which indicates that the idea of conservation in individual cases is completely general. 

Our generalisation follows from reaching a deeper understanding of the role of the preparer.  In the example discussed in \cite{boxParadoxQuestion,boxParadoxSolution}, the crucial element was the fact that at the end of the experiment the system was found with a value of conserved quantity that is far larger than any of the values it had in the original superposition.  That led to the search for its source, which turned out to be the preparer. On the other hand, that crucial element is not present here. In the situation described in our paper the particle is always found with a value of the conserved quantity that is one of the values existing in its initial superposition, hence apparently there is no need and no role for the preparer as a source of some additional amount of the conserved quantity.  Yet, as we will show, the preparer still has an essential role.

Equally important, apart from arguing that conservation laws can be extended to individual cases, we present another main result: We show that conservation can be localised at the level of the system of interest and its relevant frame of reference. Indeed, if we were not able to make sure that in a process a certain conserved quantity spreads only on a well defined set of objects, the whole issue of conservation would become unfalsifiable, and of no practical and conceptual meaning: Whenever we were presented with something that is not conserved we could claim that in fact conservation holds, but the conserved quantity is just somewhere else. This is an issue also in classical physics. If conservation of, say, momentum could only be proven by measuring the momentum of the whole universe, it would be meaningless. In classical physics, this issue is easy to be dealt with, and hence it is usually ignored. In quantum mechanics however, especially when dealing with individual cases as we do here, it becomes non-trivial, and needs to be addressed carefully. Showing that it is possible to isolate our systems, despite the propagation of entanglement and the role entanglement plays in the conservation, is another one of our main results. As a subsidiary result, this also implies that in order to understand conservation laws we don't need to make any assumptions on the state of the entire universe, such as the universe being in a state of well defined value of a conserved quantity. 

Finally we single out among all the possible interactions which conserved the quantity of interest, a particular class which we call "frame of reference type" interactions (see section \ref{sectionWhatIfNoPreparer}).  As far as statistical conservation is concerned, this particular type of conserving interactions has hitherto not been singled out as something particularly important. Yet, when considering conservation in individual cases, they have a crucial role. 

We made our arguments in the case of angular momentum of a particle on a circle, where many technicalities simplify, and bring arguments to show that this holds in full generality.

\section{Potential Solutions}

To better appreciate the problem, it is important to first look at a concrete example along with some proposed solutions (a review of the literature is given in \cite{manyWorldsConservation,jumpWhenMeasuringEnergy}, see also \cite{angularMomentumConservationPair, weakMeasurementConservation,energyMeasuringDeviceConservation}.)

The simplest example is measuring the angular momentum of a particle moving on a circle which begins in the state $\frac{1}{\sqrt{2}}(\ket{-1} + \ket{1})$, where the state $\ket{-1}$ has angular momentum $-1$, and $\ket{1}$ has angular momentum $1$.  For each individual outcome of the measurement, the total angular momentum distribution collapses to a single point, $\ket{-1}$ or $\ket{1}$.  The usual conservation of angular momentum doesn't say much: simply that the distribution of total angular momentum after the measurement is the same as the distribution calculated from the state before the measurement.  However talking in distributions is unsatisfying.  We started from a definite pure quantum state and nevertheless a quantity which is supposed to be conserved can be found, apparently at random, to be in either of two different values.  

As mentioned in the introduction, in classical mechanics the solution would be simple: the initial state is a probability distribution which represents our lack of knowledge of the true state of the system.  The measurement merely uncovers this true state, so if we find $1$ in the measurement it means that even prior to the measurement the angular momentum was $1$, and similarly for $-1$.  It's tempting to use this line of thinking in the quantum case also, and imagine that the angular momentum always had a specific value, even in the initial superposition state.  However Bell inequalities \cite{Bell,CHSH,CGLMP} show that there is no local hidden variable theory which matches quantum mechanics, by looking at the correlations we get from measuring the angular momentum in various directions on each of a pair of spin-$1/2$ particles.  Hence, when we measure the angular momentum of one of the spins and the state collapses to one of the possible values, it is {\it not} the case that we uncover a value that was already there. 

An alternative idea is that the jump in angular momentum of the particle comes from the measuring device. We shall show that this idea is also incorrect: the measuring device does not supply, receive, or catalyse any total angular momentum changes. 

One more idea is to resort to the many worlds theory, and say that the whole distribution is realised across the many worlds, and so at the level of the multiverse there is no jump \cite{manyWorldsConservation}.  However the conservation law is still written in terms of distributions, and we can only access our world in which we still see a jump, so this theory hasn't really helped us.

\section{The Measurement Process}

To proceed towards our solution, we shall look in more detail at the entire measurement process.  

To make our arguments as simple as possible, in the present paper we will focus on the case  of conservation of $\hL$, the angular momentum of a particle moving on a circle in 2-dimensional space.

What makes this example simpler than, say, conservation of linear momentum, is that angular momentum is a quantity which can be measured by itself, without any need to refer to some external system. If you are in a closed room on a rotating carousel, you can feel your head spinning without any need to look outside. This is in contrast to linear momentum: if you are in a train moving on a straight line with constant velocity, you can't know what the velocity is. Linear velocity, and by extension linear momentum, can only be defined relative to some external frame of reference, say the trees next to the rail tracks.  The measuring device needed for measuring $\hL$ does not need to refer to any frame of reference, making things simpler. 

The second simplification is that there are no other conserved quantities in the problem, so we don't have to worry that our measuring device needs to change those. For example, in 3-dimensional space, a measuring device that measures $\sigma_x$ would necessarily disturb the other spin components. 

We describe the measuring interaction by
\beq
\begin{split}
\label{measuringInteraction}
\ketNoResize{\hL= l_1}_{\vphantom{\hat{L}} S} \ket{\hq=0}_M &\rightarrow \ketNoResize{\hL=l_1}_{\vphantom{\hat{L}} S} \ket{\hq=1}_M \\
\ketNoResize{\hL=l_2}_{\vphantom{\hat{L}} S} \ket{\hq=0}_M &\rightarrow \ketNoResize{\hL=l_2}_{\vphantom{\hat{L}} S} \ket{\hq=2}_M,
\end{split}
\eeq
where $S$ represents the system to be measured, $M$ is the measuring device, and $\hq$ is its pointer which starts in the ``ready" position $\hq=0$ and at the end of the measurement indicates the measurement results.  Going forward we shall drop $\hL$ and $\hq$ in the states unless required for clarity.  Obviously this measuring interaction conserves $\hL$.

Suppose now that the system $S$ starts in the state $\alpha \ket{l_1}_S + \beta \ket{l_2}_S$, with $|\alpha|^2+|\beta|^2=1$.
The measuring interaction is then described by
\beq
\label{entangled}
 \left( \alpha \ket{l_1}_S + \beta \ket{l_2}_S \right) \ket{0}_M \rightarrow \alpha \ket{l_1}_S \ket{1}_M + \beta \ket{l_2}_S \ket{2}_M.
\eeq
Upon reading the pointer, the state collapses to 
\beq
\label{collapse}
\begin{cases}
&\ket{l_1}_S \ket{1}_M {\rm ~~~with~probability~|\alpha|^2}\\ 
&\ket{l_2}_S \ket{2}_M {\rm ~~~with~probability~|\beta|^2}.
\end{cases} 
\eeq
The question we raise is that of $\hL$ conservation during this measurement process.  There are a few basic points to be considered.

\begin{enumerate}[wide=0pt]
    \item During the time evolution in Eq. \eqref{entangled} $\hL$ is conserved in the sense that the probability distribution of $\hL$ of the system+pointer {\it calculated} for the entangled state after the measuring interaction, $\alpha \ket{l_1}_S \ket{1}_M + \beta \ket{l_2}_S \ket{2}_M$, is the same as the one {\it calculated} in the initial state of the system $\alpha \ket{l_1}_S + \beta \ket{l_2}_S$ i.e. $l_1$ with probability $|\alpha|^2$, $l_2$ with probability $|\beta|^2$.

    \item After reading the measuring device, the statistical distribution of {\it observed} values of $\hL$ is still unchanged.  Indeed this is what the usual conservation law states.

    \item Conditioned on any outcome of the measurement, the distribution collapses onto $l_1$ or $l_2$, so $\hL$ has changed. 

    \item This change in $\hL$ does not come from the measuring device, since Eq. \eqref{measuringInteraction} shows that the angular momentum of the measuring device is the same before and after the measurement: only the pointer $\hq$ changes, and this could be represented by a different type of variable such as colour.  The change in $\hL$ also does not come from the measuring interaction, nor from the act of a human observing the pointer. 
\end{enumerate}

\section{Conservation In Every Outcome} 

We now come to the crux of our paper.  Let us consider how the angular momentum superposition state has been prepared in the first place. 

Suppose we start with our particle at rest, i.e. in an eigenstate of zero angular momentum, $\ket{0}_S$, and we want to prepare, in full generality, the state 
\beq
\ket{\Psi}_S = \sum_{m=-\infty}^{\infty} \Psi(m) \ket{m}_S.
\eeq
Something needs to give some angular momentum to the particle.  We call that a ``preparer". 
  
To be able to follow angular momentum conservation, we take the preparer $P$ to also be a system moving on the circle, with initial state $\ket{\Phi}_P$, and let the interaction conserve {\it total} angular momentum:  
\beq
\label{preparedSystem}
\ket{\Phi}_P \ket{0}_S \rightarrow \sum_{m=-\infty}^{\infty} \Psi(m) \ket{\Phi - m}_P \ket{m}_S, 
\eeq
where $\ket{\Phi - m}_P$ represents a state of the preparer equal to $\ket{\Phi}_P$ shifted down in angular momentum by $m$. To anticipate the discussion in section \ref{sectionWhatIfNoPreparer}, this is not the most general total angular momentum conserving interaction, but it is an essential case for our arguments so we consider it first.

We now reached an essential point.  The above state is not exactly what we wanted - it is an entangled state between the system and the preparer.  What we wanted is to prepare the system in the pure state $\ket{\Psi}_S$.  More than that, due to angular momentum conservation the state of the system can never be pure - each component $\ket{m}_S$ will be correlated to a shifted state of the preparer.  We can however make the state of the system as close as we want to our target pure state, by making the overlap between $\ket{\Phi - m_1}_P$ and $\ket{\Phi - m_2}_P$ to be close to $1$.  This is a general property of any preparer state with narrow angular distribution (which, as we will see, also plays the role of frame of reference). 

Yet, although the state of the system becomes purer and purer, this residual entanglement with the preparer is essential.  It is still the case that $\ket{m}_S$ is correlated with $\ket{\Phi - m}_P$, and even though $\ket{\Phi - m_1}_P$ and $\ket{\Phi - m_2}_P$ become almost identical, the difference in the average value of their angular momentum does not change, remaining $m_2-m_1$.

Next, we measure $\hL$ using a generalized version of our earlier measurement interaction, Eq. \eqref{measuringInteraction}, 
\beq
\ketNoResize{\hL= m}_{\vphantom{\hat{L}} S} \ket{\hq=0}_M \rightarrow \ketNoResize{\hL=m}_{\vphantom{\hat{L}} S} \ket{\hq=m}_M.
\eeq
This gives the state:
\beq
\sum_{m=-\infty}^{\infty} \Psi(m) \ket{\Phi - m}_P \ket{m}_S
\ket{m}_M.
\eeq
Upon reading the pointer, the state collapses and we find that the system gained angular momentum $m$ and the state of the preparer was shifted down by an angular momentum of $m$.  In each case, total $\hL$ summed across the preparer and the system is conserved.

To summarise: if we look at the system alone, it is, within an approximation that we can make as accurate as we want, in the superposition state $\ket{\Psi}_S$.  Upon measuring the angular momentum, we find that in each individual case it changes, jumping to one of various possible values of $m$.  The {\it statistical} distribution is conserved, being the same as in the initial state.  However, there appears to be no conservation in each individual case. 

What we found by taking also the preparer into account is that in every single individual case, the angular momentum of the combined system+preparer is the same, and it is the same as in their initial state.  

In other words, if we look at the system alone we find the usual conundrum. There is non-conservation in individual cases, and only statistical conservation overall. If however we take into account the system+preparer, each individual case presents conservation, something that {\it is not required} by the statistical conservation law. 

\section{Who Prepared the Preparer?}
\label{whoPreparedThePreparer}

Following the above discussion, an important question arises: The preparer itself is in a coherent superposition of angular momentum states.  Who prepared the preparer?  Was it a grand-preparer?  And do we need a great grand-preparer to prepare that, and so on ad infinitum?  And, more importantly for us here, does this entire chain affect the angular momentum conservation in our experiment? The answer is no: Insofar as the conservation law is concerned, even if there is an entire chain of preparers of preparers, we only need to consider the first.  We will show that the change in our conserved quantity when a system is measured can be traced back only to the preparer and no further. 

This feature of being able to separate a small part of the world and model it in a self contained fashion is a fundamental property of nature. Everything would be completely different from how it is, if this were not so. True, one could argue that any particle in nature is affected by every other through e.g. gravity, and claim that it's impossible to model anything without modelling the entire universe.  However in practice we routinely find that we can, with an approximation as close as desired, make a self-consistent model of a small system inside the universe.  This is very important as far as conservation laws are concerned, as if we have a law which says total momentum is conserved, it is unphysical if we need to measure the momentum of the rest of the universe to see the conservation involving our particle, since the universe may be infinite, and we are unable to measure its momentum. In fact we can arrange setups such that conservation holds in a small part of the universe: in our case, the system we are considering and a preparer.  The preparer is therefore finite: it does not expand via an endless chain of grand-preparers into the rest of the universe. 

We emphasize again: this is at the core of what conservation laws really are. In particular, this avoids a trivial misinterpretation of the individual event conservation law that we described above, which says that ``the universe is in a well-defined total angular momentum $L_u$ hence whenever we find a subsystem in some state $L_s$ the rest of the universe ends in $L_u-L_s$''. This is {\it not} the content of our results, rather we claim that the conservation law in individual cases holds in a small part of the universe irrespective of what the total angular momentum of the universe is. 

We prove the separation between the preparer+system and the rest of the universe as follows.  We start with the grand-preparer, $G$, in state $\ket{\Phi_g}_G$, which has a wide spread in $\hL$ and a small spread in its conjugate variable.  The preparer, $P$, and system, $S$, begin in states with $\hL = 0$.  Consider again angular momentum preserving interactions similar to Eq. \eqref{preparedSystem}.  We first prepare the preparer into a state close to  $\ket{\Phi_p}_P$, using the $\hL$ preserving unitary operation defined for all $\ket{k}_G$ as
\beq
\label{preparePreparerUnitary}
\ket{k}_G \ket{0}_P \rightarrow \sum_{l = - \infty}^{\infty} \Phi_p(l) \ket{k-l}_G \ket{l}_P, \eeq
which gives
\beq
\label{preparePreparer}
\sum_{k = - \infty}^{\infty} \Phi_g(k) \ket{k}_G \ket{0}_P 
\rightarrow \sum_{k,l = - \infty}^{\infty} \Phi_g(k) \Phi_p(l) \ket{k-l}_G \ket{l}_P.
\eeq
Next we prepare the system in a state close to $\ket{\Psi}_S$ using 
\beq
\label{prepareSystemUnitary}
\ket{l}_P \ket{0}_S \rightarrow \sum_{m = - \infty}^{\infty} \Psi(m) \ket{l-m}_P \ket{m}_S, \eeq
which gives
\beq
\sum_{k,l,m = - \infty}^{\infty} \Phi_g(k) \Phi_p(l) \Psi(m) \ket{k-l}_G \ket{l- m}_P \ket{m}_S. 
\eeq
Suppose we measure $\hL$ on the system and get the outcome $l_0$.  The combined state of the preparer and grand-preparer is then
\beq
\label{measuredPrepareGrandpreparer}
\sum_{k,l = - \infty}^{\infty} \Phi_g(k) \Phi_p(l) \ket{k-l}_G \ket{l-l_0}_P.
\eeq

We can now calculate the distribution of the preparer and of the grand-preparer both immediately after the preparer is prepared in Eq. \eqref{preparePreparer}, and after the system is measured and found to have $\hL = l_0$.  This is done in Appendix \ref{appendixWhoPreparedThePreparer}, and the results are shown in table \ref{tablePreparerDist}.

\begin{table}[ht]
    \centering
    \begin{tabular}{c|c|c}
         & Preparer & Grand-Preparer \\
        Time & $\mathbb{P}(\hL_p = l)$ & $\mathbb{P}(\hL_g = k)$ \\
         \hline
        Preparer prep'd & $|\Phi_p(l)|^2$ & $\sum_{l = - \infty}^{\infty} |\Phi_g(k + l)|^2 |\Phi_p(l)|^2$ \\
        Measure $\hL_s = l_0$ & $|\Phi_p(l + l_0)|^2$ & $\sum_{l = - \infty}^{\infty} |\Phi_g(k + l)|^2 |\Phi_p(l)|^2$ \\
    \end{tabular}
    \caption{Distribution of $\hL$ for the Preparer and Grand-Preparer after the Preparer is prepared, and after $\hL$ is measured on the System and found to be $l_0$.}
    \label{tablePreparerDist}
\end{table}

This shows that the distribution of the angular momentum of the preparer changes by $-l_0$, offsetting the change in the system. On the other hand, the distribution of angular momentum of the grand-preparer is the same at these two times.  The grand-preparer gave some of its angular momentum to the preparer for its initial state, but that's all. 

To conclude, although the slight residual entanglement between the system and preparer was crucial, the residual entanglement between the preparer and the grand-preparer is irrelevant. Hence we have no need to include any other part of the universe outside our system and preparer in order to see that conservation is maintained in each individual outcome. This is essential: to understand the conservation laws under measurement we don't need to make any cosmologic assumptions about the angular momentum distribution of the universe. 

\section{What if there is no Preparer?  Frames of Reference}
\label{sectionWhatIfNoPreparer}

Thus far we argued that the way to see conservation in every measurement outcome is to include the preparer.  However one may ask, what happens if someone gives you a system which is in precisely the superposition state of a conserved quantity $\alpha \ket{l_1} + \beta \ket{l_2}$.  How can conservation be maintained if you measure angular momentum on the state and find $\ket{l_1}$, when the state has no entanglement with any preparer?

A related question concerns our preparation interaction, both for the preparer-system and the grand preparer-preparer. The interaction considered in Eqs. (\ref{preparedSystem},\ref{preparePreparerUnitary},\ref{prepareSystemUnitary}) is {\it not} the only one possible. Indeed, in order to conserve total angular momentum the interaction can depend on the relative system-preparer angle or on the angular momentum of the system or of the preparer, or on a combination of these.  It only can not depend on the sum of their absolute angles.  Considering more general procedures may apparently invalidate our conclusion. For example, starting with the preparer in state $|\Phi\ra_P=\alpha|l_1\ra_P+\beta|l_2\ra_P$ and exchanging the preparer and system states gives
\beq \label{different interaction}
\left( \alpha \ket{l_1}_P + \beta \ket{l_2}_P \right) \ket{0}_S \rightarrow \ket{0}_P \left( \alpha \ket{l_1}_S + \beta \ket{l_2}_S \right),
\eeq 
which is different from our basic interaction, Eq. \eqref{preparedSystem}.
How can conservation hold when the system is not entangled with anything?

The answer to both these questions, as explained in \cite{boxParadoxSolution}, is that no-one can give you the state $\alpha \ket{l_1} + \beta \ket{l_2}$ in the first place.  The issue is: what is the meaning of the relative phase between $\ket{l_1}$ and $\ket{l_2}$?  

A superposition of angular momentum states has a non-uniform angular distribution on the circle. But the same angular distribution could be placed in a different location, rotated around the circle from the first, by appropriately changing the phases.  Thus the phase is related to the angle.  However an absolute angle is meaningless: angle only has meaning relative to a frame of reference.  In other words a state $\Psi(\theta_s)$, where $\theta_s$ is the angle describing the location of the system, has no meaning.  Instead one can only prepare $\Psi(\theta_s - \theta_f)$ where $\theta_f$ is the angle of a frame of reference.  As we show in Appendix \ref{appendixFrameReferencePreparer}, the particular preparation procedure we considered in Eq. \eqref{preparedSystem} is precisely one in which the desired state of the system is prepared {\it relative} to the location of the preparer, i.e. the preparer plays the role of a frame of reference \cite{wigner,superSelectionRules,arakiYanase,frameOfReference,aharonovFramesOfReference,frameOfReferenceReview}. Indeed, Fourier transforming Eq. \eqref{preparedSystem} into the angular representation gives, as seen in Eq. \eqref{prepareSystemAngular},
\beq
\label{prepareSystemAngularMain}
\begin{split}
&\int \tilde{\Phi}_p(\theta_p) \ket{\theta_p}_P d \theta_p \ketNoResize{\hL_s = 0}_{\vphantom{\hat{L}} S} \\
&\rightarrow \int \tilde{\Phi}_p(\theta_p) \ket{\theta_p}_P \int \tilde{\Psi}(\theta_s - \theta_p) \ket{\theta_s}_S d\theta_s d\theta_p,
\end{split}
\eeq
where $\theta_p$ is the angle of the preparer, which here acts as a frame of reference.  We call this a ``reference-frame type interaction". Similarly, by Eq. \eqref{preparePreparer}, the grand-preparer is a frame of reference for the preparer.  

Without a preparer that plays the role of a frame of reference, the state has no meaning.  We thus postulate that a reference-frame type interaction of the form Eq. \eqref{prepareSystemAngularMain}, or equivalently Eq. \eqref{preparedSystem}, must occur.  This might not be in the first preparer-system interaction, in which case some grand-preparers down the chain will participate in the conservation process (in the example Eq. \eqref{different interaction} the preparation of the preparer requires basically the same angular momentum transfer as the preparation of the system).  Eventually however, a reference-frame type interaction will occur, and this will terminate the participations of grand-preparers. 

One may also ask whether we can find a preparer, a coherent superposition of a conserved quantity, in nature. In fact such states are perfectly natural: we have many objects which can be used as a frame of reference for an angle, and any such frame of reference can be used as a preparer for a superposition of the corresponding angular momentum.  While preparing the state with a relative angle to this frame, the exchange of angular momentum seen in Eq. \eqref{preparedSystem} occurs.  This is the case even if the universe as a whole has a definite value for the conserved quantity: we can always find a subsystem to use as our preparer.

\section{Conclusion}

We have shown that conservation laws apply in quantum mechanics for every measurement outcome, not just as a distribution.  The key to this is that even though in quantum mechanics the system is usually described by a pure state, something which is unentangled with anything else, in reality it is not so, and we need to include an extra object in our model - the preparer - with which the system is (even if only slightly) entangled \cite{boxParadoxSolution}.  When we model a system in quantum mechanics, we simplify our model as much as possible, and only keep the essential parts of the whole experiment.  Historically the preparer has been discarded after preparation of the initial state of the system. This did not ruin the conservation laws at statistical level, but gave the impression that that is all one can expect in quantum mechanics. We however have argued that the preparer is always there in the background, and shown that it should be included; the small amount of entanglement that it retains with the system has the effect of ensuring conservation in each individual case. 

We also showed that there is no need to look at the rest of the universe to see the offsetting angular momentum change when we measure one small system: we need only include the preparer.  Although this is rarely appreciated, this possibility of separating a few systems and having the conservation law apply to them is implicitly assumed in formulating a physically observable conservation law, both in classical and quantum mechanics.  Indeed, if one would have to check the entire universe to prove that a certain quantity is conserved, the conservation law would made no physical sense.  Since conservation in individual cases crucially depends on extremely small amounts of entanglement one might have thought this impossible.  We however showed that it is possible to separate systems whilst maintaining this stronger conservation law. 

Another facet of the above is related to the involvement of frames of reference needed for preparing superpositions of states with different values of a conserved quantity; the "preparer" is actually such a frame.  We then raised the question of the possible need to consider an infinite regression chain of "frames-of-reference of frames-of-reference". We showed that the conserved quantity can be kept not propagating down the chain, but be maintained within the system of interest and its relevant frame of reference; this is the key element that allows the conservation to be separated from the whole universe.

The arguments in this paper were presented for the angular momentum of a particle moving on a circle. This is arguably the simplest example since measuring it does not affect any other conserved quantities.  However in general we may have a system with more conserved quantities which do not commute with each other, so measuring one disturbs another, e.g. angular momentum in a particular direction in three-dimensional space does not commute with angular momentum in any other direction.  We believe that conservation laws hold for every measurement outcome for these conserved quantities also.  Specifically, we conjecture that when measuring the angular momentum along the $z$-axis, as in the simple example described in our paper, we can do it without the measuring device providing the required angular momentum along the $z$-axis.  Instead the preparer provides that $z$ angular momentum.  At the same time, the measuring device disturbs the $x$ and $y$ angular momenta of the system: for those components it is the measuring device which exchanges angular momentum with the system ensuring conservation.  An analysis of such cases is reserved for future work.

Thus we argue that the conservation laws in quantum mechanics need to be restated, to say that the total of a conserved quantity, summed across all systems including the preparer, is unchanged at all times and in each individual measurement outcome.

Finally, zooming out, the results we present here sit on one end of a larger spectrum. We proved, with all the caveats, that conservation laws work at the {\it individual} level in what is arguably the most fundamental, yet simplest case - when a superposition of states with different values of a conserved quantity is prepared and then directly subjected to a measurement of the conserved quantity. The preparer plays an essential role, but the way in which it works is ultimately straightforward.  At the other end of the spectrum, if between the preparation and measurement the system is subjected to some other interaction, the way conservation is maintained is more subtle, and contains a few more basic ingredients \cite{boxParadoxSolution}.

We expect that the processes discovered in \cite{boxParadoxSolution}, together with those discussed here, will lead us to establish that quantum mechanical conservation laws work at the individual case level, as opposed to merely the statistical level, in any physical situation.

\section{Acknowledgements}
We thank Tony Short, Paul Skrzypczyk, Yakir Aharonov and Carolina Moreira Ferrera for helpful discussions and advice.  We are supported by the European Research Council Advanced Grant FLQuant.

\bibliographystyle{quantum}
\bibliography{ConservationLaws}

\appendix

\section{Who Prepared The Preparer? Frames of Reference}
\label{appendixWhoPreparedThePreparer}

In this appendix we give the calculations for section \ref{whoPreparedThePreparer}, 
which analyses the way conservation works for a system prepared by a preparer which in turn was prepared by a grand-preparer.  We will calculate the distribution of the angular momentum $\hL$ for both the preparer and the grand-preparer.  We calculate it first after the grand-preparer has prepared the preparer but before the system is prepared, and second after the system is prepared and measured and found to have a value $l_0$.  The results of these calculations shows that the distribution of the grand-preparer is unchanged, whereas the distribution of the preparer is shifted down by $l_0$.

Consider first the preparation of the preparer by the grand preparer. Their state before the preparer preparing the system is  
\beq
\ket{\kappa}_{G,P} = \sum_{k,l = - \infty}^{\infty} \Phi_g(k) \Phi_p(l) \ket{k-l}_G \ket{l}_P,
\eeq
as described in Eq. \eqref{preparePreparer}.

At this time the distribution of the angular momentum of the preparer, $\hL_p$, is given by
\beq
\begin{split}
&\mathbb{P}(\hL_p = l) = \leftindex[][|]_{G,P}{\la} \kappa| \ket{l}_P \leftindex_P{\bra{l}} \ket{\kappa}_{G,P} \\
&= \sum_{k,k' = - \infty}^{\infty} \Phi^*_g(k')\Phi_g(k) | \Phi_p(l) |^2 \leftindex_G{\braket{k'-l}{k-l}}_G \\
&=\sum_{k,k' = - \infty}^{\infty} \Phi^*_g(k')\Phi_g(k) | \Phi_p(l) |^2 \delta_{k',k} \\
&= | \Phi_p(l) |^2.
\end{split}
\eeq
This is the same as the distribution for $\hL_p$ in the state $\Phi_p$, i.e. in the preparer state we were trying to prepare.  The distribution for the angular momentum of the grand-preparer, $\hL_g$, at this time is given by 
\beq
\label{grandPreparerInitialDistn}
\begin{split}
&\mathbb{P}(\hL_g = k)  \\
&= \leftindex[][|]_{G,P}{\la} \kappa| \ket{k}_G \leftindex_G{\bra{k}} \ket{\kappa}_{G,P} \\
&= \sum_{k'',k',l,l' = - \infty}^{\infty} \Phi^*_g(k')\Phi_g(k'') \Phi^*_p(l') \Phi_p(l) \times \\ 
& \qquad \qquad \qquad \qquad \qquad  \leftindex_G{\braket{k'-l'}{k}}_G \leftindex_G{\braket{k}{k''-l}}_G \leftindex_P{\braket{l'}{l}}_P \\
&= \sum_{k'',k',l = - \infty}^{\infty} \Phi^*_g(k')\Phi_g(k'') \Phi^*_p(l) \Phi_p(l) \delta_{k' - l,k} \delta_{k,k'' - l} \\
&= \sum_{l = - \infty}^{\infty} | \Phi_g(k + l) |^2 | \Phi_p(l) |^2,
\end{split}
\eeq
i.e. it is the average over the distribution of the preparer, $\Phi_p(l)$, of the initial distribution of the grand-preparer shifted down by $l$, $\Phi_g(k + l)$.

Let now the preparer prepare the system, and then measure the angular momentum of the system. If the system is found with angular momentum $l_0$, the state of the preparer and grand-preparer is 
\beq
\ket{\chi}_{G,P} = \sum_{k,l = - \infty}^{\infty} \Phi_g(k) \Phi_p(l) \ket{k-l}_G \ket{l-l_0}_P,
\eeq
as given in Eq. \eqref{measuredPrepareGrandpreparer}.  The angular momentum distributions of the preparer is   
\beq
\begin{split}
&\mathbb{P}(\hL_p = l | \hL_s = l_0) \\
&= \leftindex[][|]_{G,P}{\la} \chi| \ket{l}_P \leftindex_P{\bra{l}} \ket{\chi}_{G,P} \\
&= \sum_{k,k',l'',l' = - \infty}^{\infty} \Phi^*_g(k')\Phi_g(k)\Phi^*_p(l')\Phi_p(l'') \\
& \qquad \qquad \qquad \cdot \leftindex_G{\braket{k'-l'}{k-l''}}_G \leftindex_P{\braket{l'-l_0}{l}}_P \leftindex_P{\braket{l}{l''-l_0}}_P\\
&= \sum_{k,k' = - \infty}^{\infty} \Phi^*_g(k')\Phi_g(k) |\Phi_p(l + l_0)|^2 
\delta_{k' - (l + l_0),k - (l + l_0)}\\
&= |\Phi_p(l + l_0)|^2.
\end{split}
\eeq
i.e. it's the distribution of the preparer in its initial state, $|\Phi_p(l)|^2$, shifted down by the angular momentum of the system, $l_0$.
This does not depend on the grand preparer at all, and hence is the same distribution as we would get if we had only a preparer and no grand-preparer.

Similarly, the distribution of the angular momentum of the grand-preparer, $\hL_g$, is now 
\beq
\begin{split}
&\mathbb{P}(\hL_g = k | \hL_s = l_0) \\
&= \leftindex_{G,P}{\bra{\chi}} \ket{k}_G \leftindex_G{\bra{k}} \ket{\chi}_{G,P} \\
&= \sum_{k'',k',l,l' = - \infty}^{\infty} \Phi^*_g(k')\Phi_g(k'')\Phi^*_p(l')\Phi_p(l) \\
& \qquad \qquad \qquad \cdot \leftindex_G{\braket{k'-l'}{k}}_G \leftindex_G{\braket{k}{k''-l}}_G \leftindex_P{\braket{l'-l_0}{l-l_0}}_P \\
&= \sum_{l = - \infty}^{\infty} |\Phi_g(k+l)|^2 |\Phi_p(l)|^2.
\end{split}
\eeq
This is the same distribution for the grand-preparer as we had prior to preparing the system, in Eq. \eqref{grandPreparerInitialDistn}.  Thus the preparation and measurement of the system has no effect on the distribution of the grand-preparer.  We have no need to include the grand-preparer in our models. 

\section{Frame of Reference as a Preparer}
\label{appendixFrameReferencePreparer}

Here we shall show that the frame of reference required to define an angle is the same object as the preparer we used to create a superposition state of angular momentum.

Suppose we wish to prepare a system $S$ in the state
\beq
\begin{split}
\ket{\Psi}_S 
&= \sum_{m = - \infty}^{\infty} \Psi(m) \ketNoResize{\hL_s = m}_{\vphantom{\hat{L}} S} \\
&= \int \tilde{\Psi}(\theta_s) \ketNoResize{\htheta_s = \theta_s}_{\vphantom{\hat{L}} S} \, d\theta_s, \\
\end{split}
\eeq
where $\hL_s$ represents the conserved quantity angular momentum, and $\htheta_s$ the angle.  However the absolute angle $\theta_s$ has no meaning; what we actually want is to prepare the state {\it relative} to a frame $F$, i.e. to prepare 
\beq
 \ket{\theta_f}_F \int \tilde{\Psi}(\theta_s - \theta_f) \ket{\theta_s}_S d\theta_s. 
\eeq
where $\theta_f$ is the location of the frame. In general, the frame doesn't have a perfectly well defined location (which would also be a non-normalisable state); we take it as a narrow angular distribution 
 $\theta_f$,
\beq
\ket{\Phi_f}_F = \int \tilde{\Phi}_f(\theta_f) \ket{\theta_f}_F d\theta_f,
\eeq
leading to
\beq
\label{refPreparedState}
\int \tilde{\Phi}_f(\theta_f) \ket{\theta_f}_F \int \tilde{\Psi}(\theta_s - \theta_f) \ket{\theta_s}_S d\theta_s d\theta_f. 
\eeq
 
We wish to compare this state to that created from a preparer $P$, which starts in a state 
\beq
\ket{\Phi_p}_P = \sum_{l = - \infty}^{\infty} \Phi_p(l) \ket{l}_P. 
\eeq
We prepare $S$ using the unitary transformation defined for all $l$ as
\beq
\ket{l}_P \ketNoResize{\hL_s = 0}_{\vphantom{\hat{L}} S}  \rightarrow \sum_{m = - \infty}^{\infty} \Psi(m) \ket{l - m}_P \ket{m}_S,
\eeq
which gives
\beq
\ket{\Phi_p}_P \ketNoResize{\hL_s = 0}_{\vphantom{\hat{L}} S} \rightarrow \sum_{l = - \infty}^{\infty} \Phi_p(l) \sum_{m = - \infty}^{\infty} \Psi(m) \ket{l - m}_P \ket{m}_S.
\eeq

If we look at this preparation in the angular representation, applied on an angle basis vector, we get:
\beq
\begin{split}
&\ket{\theta_p}_P \ketNoResize{\hL_s = 0}_{\vphantom{\hat{L}} S} \\
&= \sum_{l = - \infty}^{\infty} e^{- i \theta_p l} \ket{l}_P \ketNoResize{\hL_s = 0}_{\vphantom{\hat{L}} S}  \\
&\rightarrow \sum_{l = - \infty}^{\infty} e^{- i \theta_p l} \sum_{m = - \infty}^{\infty} \Psi(m) \ket{l - m}_P \ket{m}_S\\
&= \sum_{l,m = - \infty}^{\infty} e^{-i \theta_p (l - m)} e^{-i \theta_p m} \Psi(m) \ket{l - m}_P \ket{m}_S \\
&= \sum_{m = - \infty}^{\infty} e^{-i \theta_p m} \Psi(m) \ket{\theta_p}_P \ket{m}_S\\
&= e^{-i \theta_p \hL_s} \ket{\theta_p}_P \sum_{m = - \infty}^{\infty} \Psi(m) \ket{m}_S\\
&= e^{-i \theta_p \hL_s} \ket{\theta_p}_P \int \tilde{\Psi}(\theta_s) \ket{\theta_s}_S d\theta_s \\
&= \ket{\theta_p}_P \int \tilde{\Psi}(\theta_s) \ket{\theta_s + \theta_p}_S d\theta_s, \\
&= \ket{\theta_p}_P \int \tilde{\Psi}(\theta_s - \theta_p) \ket{\theta_s}_S d\theta_s.
\end{split}
\eeq
Therefore on the initial state of the preparer the preparation does
\beq
\label{prepareSystemAngular}
\begin{split}
&\int \tilde{\Phi}_p(\theta_p) \ket{\theta_p}_P d \theta_p \ketNoResize{\hL_s = 0}_{\vphantom{\hat{L}} S} \\
&\rightarrow \int \tilde{\Phi}_p(\theta_p) \ket{\theta_p}_P \int \tilde{\Psi}(\theta_s - \theta_p) \ket{\theta_s}_S d\theta_s d\theta_p.
\end{split}
\eeq

Taking $P$ as $F$ and $\Phi_p = \Phi_f$, we see we have prepared the state $\ket{\Psi}_S$ relative to the frame of reference, Eq. \eqref{refPreparedState}.  The preparer {\it is} the frame of reference.

\end{document}